\documentstyle[12pt,aps,prc,preprint,epsfig]{revtex}
\tightenlines

\begin{document}

\title{Semiempirical Shell Model Masses with Magic Number Z = 126 for
Superheavy Elements}
\author{S. Liran\footnote{Present address: Kashtan 3/3, Haifa 34984,
Israel}, A. Marinov and N. Zeldes \\
The Racah Institute of Physics, The Hebrew University of Jerusalem, Jerusalem 91904,
Israel}
\maketitle

\begin{abstract}
A semiempirical shell model mass equation applicable to superheavy elements
up to Z = 126 is presented and shown to have a high predictive power.
The equation is applied to the recently discovered superheavy nuclei
$^{293}118$ and $^{289}114$ and their decay products.
\end{abstract}


\bigskip

Recently \cite{nin99} an $\alpha$-decay chain consistent with the
formation of ${^{293}118}$ and its sequential decay to ${^{289}116},
{^{285}114}, {^{281}112}, {^{277}110}, {^{273}{\rm Hs}}$ and ${^{269}{\rm
Sg}}$ has been observed. The $\alpha$-decay energies vary rather
smoothly along the chain. If the above assignments are confirmed, and
the decaying nuclei are formed in or near their ground states (g.s.)
\cite{cnh99}, then the smooth variation seems to preclude the
traditional macroscopic-microscopic \cite{mon94,smo97} Z = 114 as
a magic proton number in these nuclei.

Recent phenomenological studies of BE(2) systematics \cite{zam95}
and of the persistence of the Wigner term in masses of heavy
nuclei \cite{zel98} indicate Z = 126 as the next spherical proton
magic number after lead, and this is consistent with
considerations based on nuclear diffuseness \cite{mys98}. Recent
self-consistent and relativistic mean field calculations
\cite{cwi96,rut97,ben99,kru00} variously predict proton magicities
for Z = 114, 120, 124 and 126.

These new developments are contrary to the assumption made in the
semiempirical shell-model mass equation (SSME) \cite{liz76} (see also
ref. \cite{zel96}) that Z = 114 is the next proton magic number beyond
lead. The equation stops at Z = 114, and it is unsuitable for
extrapolation already earlier, beyond Hs (Z = 108), as shown by its
increasing deviations from the data beyond that (like in fig. 4 of
ref. \cite{nin99}). One has to find an appropriate substitute for
the equation in the neighbourhood of Z = 114 and beyond.

During the early stages of the SSME \cite{liran}, when it was
adjusted separately in individual shell regions in the N $-$ Z plane,
both Z = 114 and Z = 126, which were at the time considered possible
candidates for the post-lead proton magic number, were tried as a shell
region boundary in each of the two heaviest regions with Z $\geq$ 82
and respective N boundaries 82~$\leq$~N~$\leq$~126 (called here region A)
and 126~$\leq$~N~$\leq$184 (called region B). The agreement with the data
was about the same for both choices, and considering the prevailing
view in the mid nineteen seventies Z = 114 was chosen for the SSME
mass table. In this Letter we consider the possibility of substituting
the SSME \cite{liz76} in region B with the early results obtained with
Z = 126. In particular we study the predictive power or
extrapolatability of these results by using the newer data
accumulated after the adjustments were made, like in refs.
\cite{hau84,mol95,mnk97,abo95}. Then we apply the equation to the
results of two recent superheavy elements (SHE) experiments
\cite{nin99,oga99}. Region A will be considered elsewhere.

In the SSME the total nuclear energy E is written as a sum of
pairing, deformation and Coulomb energies:

\begin{equation}
E\left( {N,Z} \right)=E_{pair}\left( {N,Z} \right)+E_{def}\left( {N,Z}
\right)+E_{Coul}\left( {N,Z} \right)\ .
\label{eq1}
\end{equation}

The form of $E_{Coul}$ is the same in all shell regions:

\begin{equation}
E_{Coul}\left( {N,Z} \right)=\left( {{{2Z_0} \over A}}
\right)^{1/3} [{\alpha ^C+\beta ^C\left(
{Z-Z_0} \right)+\gamma ^C\left( {Z-Z_0} \right)^2}]\  ,
\label{eq2}
\end{equation}
and that of $E_{pair}$ is the same separately in all diagonal shell
regions, where the major valence shells are the same for neutrons and
protons, and in all non-diagonal regions, where the neutron and proton
valence shells are different. In a non-diagonal region like B it is

\begin{eqnarray}
    E_{pair}(N,Z) = \left({{A_{0}}\over{A}}\right)   & & [ \alpha + \beta_{1} (N -
    N_{0}) + \beta_{2}(Z - Z_{0})  +   \nonumber \\
     & & \gamma_{1} (N - N_{0})^{2} + \gamma_{2}(Z - Z_{0})^{2}+
     \gamma_{3}(N-N_{0})(Z- Z_{0}) + \nonumber \\
     &  &  {{1 - (-1)^{N}}\over{2}}\Theta_{1} +{{1 -
     (-1)^{Z}}\over{2}}\Theta_{2} +{{1 - (-1)^{NZ}}\over{2}}\mu  ] \ .
    \label{eq3}
\end{eqnarray}
The part $E_{def}$ for region B with Z = 126 as upper proton boundary
is \cite{liran}

\begin{equation}
    E_{def}(N,Z)=\left({{A_{0}}\over{A}}\right)\left[\varphi_{21}\Phi_{21}(N,Z)+
    \varphi_{31}\Phi_{31}(N,Z)+\chi_{12}X_{12}(N,Z)\right] \ ,
    \label{eq4}
\end{equation}
with
\begin{equation}
    \Phi_{21}(N,Z) = (N - 126)^{2}(184 - N)^{2}(Z - 82)(126 - Z)
    \label{eq5}
\end{equation}

\begin{equation}
    \Phi_{31}(N,Z) = (N - 126)^{3}(184 - N)^{3}(Z - 82)(126 - Z)
    \label{eq6}
\end{equation}

\begin{equation}
    X_{12}(N,Z) = (N - 126)(184 - N)(N - 155)(Z - 82)^{2}(126 -
    Z)^{2}(Z - 104) \ .
    \label{eq7}
\end{equation}

The respective values of $N_{0}, Z_{0}$ and $A_{0}$ are 126, 82 and
208. The coefficients multiplying the functions of N and Z are
adjustable parameters which were determined by a least-squares
adjustment to the data \cite{liran}. Their numerical values are given
in table I. The mass excesses $\Delta M(N,Z)$ are obtained by adding
to eq.~(\ref{eq1}) the sum of nucleon mass excesses $N\Delta M_{n}+
Z\Delta M_{H}$.

The experimental data used in the adjustments included 211 masses.
(Ref. \cite{wag71} augmented by data from the literature up to Spring
1973.) Presently there are 267 known experimental masses in region B.
(Ref. \cite{auw95} (excluding values denoted ``systematics'' (\#))
augmented by data from the literature.) They include 56 new masses that
were not used in the adjustments.

Unlike the SSME \cite{liz76}, when the corresponding 56 deviations
of the predictions of eq.~(\ref{eq1}) (with the definitions
(\ref{eq2})-(\ref{eq4})) from the data are plotted as function of
Z they do not increase when Z increases towards 114. On the other
hand, there are conspicuously large deviations of $^{218,219}$U
and also $^{217,219}$Pa, with respective neutron numbers 126, 127,
126 and 128.

Fig. 1 shows the deviations as function of the distance from the
line of $\beta$-stability, denoted ``neutrons from stability''
(NFS) and defined by NFS = $N - Z - \frac{0.4A^{2}}{A + 200}$
\cite{hau84}. Empty circles denote the deviations of the $N =
126-128$ nuclei $^{216}$Ac, $^{218}$Pa, $^{216}$Th, $^{217}$Pa,
$^{219}$Pa, $^{219}$U and $^{218}$U, which increase in this order.
These deviations indicate increasing underbinding of extrapolated
$N \approx 126$ nuclei when Z increases. They are related to the
increasing discontinuity of the extrapolated mass surface along
the common boundary  N = 126 of regions A and B away from the
data, when the two regions are adjusted separately
\cite{liran,zel67,com70}. (Such deviations can be avoided by
adjusting the data in the two regions simultaneously, with
continuity requirements along the boundary imposed as additional
constraints \cite{liz76}.)

The deviations of the remaining 49 nuclei with $N \geq 129$, which do
not follow the $N = 126$ boundary but extend into the interior of the
shell region, are marked by full circles. They are about equally
positive and negative, have similar magnitudes, and do not seem to be
correlated with NFS.

Table II, patterned after similar more elaborate ones
\cite{mol95,mnk97}, shows $\delta_{av}$ and $\delta_{rms}$, the
respective average and rms deviations of eq.~(\ref{eq1})
from the data, for $\Delta M, S_{n}, S_{p}, Q_{\beta^{-}}$ and
$Q_{\alpha}$. The deviations are shown separately for the older data
that were used in the adjustments and for the newer data. The last
column shows the error ratios $\delta_{rms}^{new}:\delta^{old}_{rms}$.

For the old data the magnitudes of the $\delta_{av}$ are single keVs,
and for the $\delta_{rms}$ they are in the range 110$-$170 keV. For the
new data they are larger, with respective highest values of 53 and
236 keV for $\Delta M$ and smaller values for $S_{n}, S_{p},
Q_{\beta^{-}}$ and $Q_{\alpha}$.

The table shows as well in brackets the corresponding deviations for
the 49 $N \geq 129$ nuclei extending into the interior of the shell
region, where SHE are presently searched for. Except for $Q_{\alpha}$
they are smaller than the unbracketed deviations.

The deviations shown in table II are about one half of the
corresponding deviations for several recent mass models
\cite{mol95,mnk97,abo95,mys96,duz95}. The main reason for these
smaller deviations is presumably the inclusion in eq.~(\ref{eq1}) of
the particle-hole-symmetric configuration interaction terms
$E_{def}$ (eq.~(\ref{eq4})). (Without these terms the rms deviation
from the original data of the part $E_{pair} + E _{Coul}$ (eqs.
(\ref{eq3}) and (\ref{eq2}) alone) is 1076 keV \cite{liran}, as
compared to the rms deviation of 126 keV from the original $\Delta M$
data in table II for the complete eq.~(\ref{eq1}). See also ref.
\cite{zgl65}.)

Until a new SSME adjustment to the data in both regions A and B is
undertaken we propose the use of eq.~(\ref{eq1}) with the coefficients
of table I instead of the SSME \cite{liz76} as an appropriate predictive
tool in SHE research in the interior of region B. It is important to
emphasize, though, that the above rather suggestive results are not
a proof of superior magicity of Z = 126 as compared to other
recently proposed predictions \cite{rut97,ben99,kru00}, because no
comparative mass studies of this kind were made. These are beyond the scope
of the present work.

We now apply the equation to the results obtained in \cite{nin99}.
Panel a of fig. 2 shows the chain of $\alpha$-decay energies measured
in \cite{nin99}, and the values predicted for them by
eq.~(\ref{eq1}) when the data are interpreted as g.s.
transitions of the nuclei assigned in \cite{nin99}. (In ref. \cite{cnh99}
some alternative possibilities of transitions between low-lying
Nilsson levels are considered.) The figure shows as well the
predictions \cite{smo99} which motivated the search undertaken in
\cite{nin99}. The respective average and rms deviations of the
predicted values from the data are $-197$ and 308 keV for
eq.~(\ref{eq1}) and $-154$ and 357 keV for ref. \cite{smo99}.
The rms deviation of eq.~(\ref{eq1}) is consistent with
table II, but the average deviation is too negative.

The variation of the predicted values of eq.~(\ref{eq1})
along the chain is smoother than that of the data. The kinks in the
data at Z = 112 and 116 are not reproduced. Such kinks are usually
interpreted as submagic number effects, and in the SSME these are
assumed to have been obliterated by configuration interaction between
subshells, described by the terms $E_{def}$, eq.~(\ref{eq4}). The
inadequacy of the SSME to describe abrupt local changes associated
with subshell structure is detailed in figs. 2a-c of ref. \cite{liz76}.

On the other hand, the microscopic energies calculated in ref.
\cite{smo99} are basically sums of (bunched minus unbunched) single
nucleon energies, and as such have (magic and) submagic gap effects
built in. The corresponding predicted line in fig. 2 has kinks at Z =
110 and 116, corresponding to the predicted submagic numbers Z = 108
and 116 indicated by increasing vertical distances between isotopic
$Q_{\alpha}$ lines in fig. 4 of \cite{smo97}.

Most of the smoothing effect of configuration interaction is missing in
macroscopic-microscopic Strutinsky type and in self-consistent mean
field calculations. The included $T=1, J=0$ pairing correlations
seem not to be enough. This might result in calculated submagic gaps and
associated kinks which are too large compared to the data. Panel b of
fig. 2 shows large kinks at respective proton numbers Z = 112, 114 and
116, predicted by refs. \cite{abo95}, \cite{mnk97} and \cite{cnh99}.
The kinks at Z = 114 and 116 were observed before \cite{nin99,ben00}.

Finally we mention the $\alpha$-decay chain observed in ref.
\cite{oga99}, which is considered a good candidate for originating
from $^{289}$114 and its sequential decay to $^{285}$112 and
$^{281}$110. The respective average and rms deviations of the
predictions of eq.~(\ref{eq1}) from the measured energies
are 847 and 905 keV, which considerably exceed the deviations expected
from table II for g.s. transitions. If the above assignments are
confirmed, the large deviations might indicate that the decay chain
does not go through levels in the vicinity of the g.s.

It might also be worthwhile mentioning that for the conceivable parents
$^{288}$112 or $^{291}$113 which can be obtained from the compound nucleus
$^{292}$114 by respective $1 \alpha$ or $1p$
evaporation, the corresponding average and rms deviations of
eq.~(\ref{eq1}) from the measured energies are $-181$ and 366 keV
and $-242$ and 417 keV, which are more than twice smaller than for the
parent $^{289}$114.

We thank Stelian Gelberg and Dietmar Kolb for help with the calculations.

\newpage

\begin{table}
    \caption{Values of the coefficients of
    eq.~(1) determined by adjustment to the data [14]}
    \begin{tabular}{crcc}
        Coefficient & Value (keV) &&  \\
        \hline
        $\alpha$ & $-2.3859605 \times 10^{6}$  &&\\
        $\beta_{1}$ & $-1.496441 \times 10^{4}$ && \\
        $\beta_{2}$ & $-3.3866255 \times 10^{4}$  &&\\
        $\gamma_{1}$ & $3.022233 \times 10^{1}$ && \\
        $\gamma_{2}$ &  $2.811903 \times 10^{1}$ && \\
        $\gamma_{3}$ &  $-3.6159266 \times 10^{2}$ && \\

        $\Theta_{1}$ & $8.16 \times 10^{2}$ &&  \\
        $\Theta_{2}$ & $1.007 \times 10^{3}$ &&  \\
        $\mu$ & $-2.121 \times 10^{2}$ && \\
        $\alpha^{C}$ & $8.111517 \times 10^{5}$ &&  \\
        $\beta^{C}$ & $2.0282913 \times 10^{4}$ &&  \\
        $\gamma^{C}$ & $1.0930065 \times 10^{2}$ &&  \\
        $\varphi_{21}$ & $-9.87874 \times 10^{-5}$ &&  \\

        $\varphi_{31}$ & $3.13824 \times 10^{-8}$ && \\
        $\chi_{12}$ & $-1.428529 \times 10^{-7}$ &&  \\
    \end{tabular}
\end{table}

\begin{table}
    \caption{Numbers of data N, average deviations $\delta_{av}$, and
    rms deviations $\delta_{rms}$, for eq.~(1) with the coefficients
    of table I. The numbers in brackets are obtained when nuclei with
    $N = 126-128$ are excluded. The last column shows the error ratios
    $\delta^{new}_{rms}:\delta^{old}_{rms}$.}
    \begin{tabular}{lcccccccc}
    & \multicolumn{3}{c}{Original nuclei (1973)} & &
    \multicolumn{3}{c}{New nuclei (1973-1999)}  \\
         &  & $\delta_{av}$ & $\delta_{rms}$ & & & $\delta_{av}$ &
         $\delta_{rms}$ & Error  \\
        Data & N & (keV) & (keV) & & N & (keV) & (keV) & ratio \\
        \hline
        $\Delta M$ & 211 & ~~2 & 126 & & 56 (49) & 53 ($-1$) & 236
(155) &
        1.87 (1.23) \\
        $S_{n}$ & 169 & ~~1 & 117 &  & 45 (38) & 12 $(-2)$ & 171
(145) &
        1.46 (1.24) \\
        $S_{p}$ & 162 & $-4$ & 121 &  & 52 (44) & $-17$ (15) & 184
(148) &
        1.52 (1.22) \\
        $Q_{\beta^{-}}$ & 146 & $-7$ & 158 &  & 51 (44) & $-19$
(14) & 209
        (169) & 1.32 (1.07) \\
        $Q_{\alpha}$ & 174 & $-6$ & 162 &  & 57 (55) & $-3~(-8)$ & 220
        (220) & 1.36 (1.36) \\
    \end{tabular}
    \protect\label{}
\end{table}
\newpage

\begin{figure}
\begin{center}
\leavevmode \epsfysize=5.0cm \epsfbox{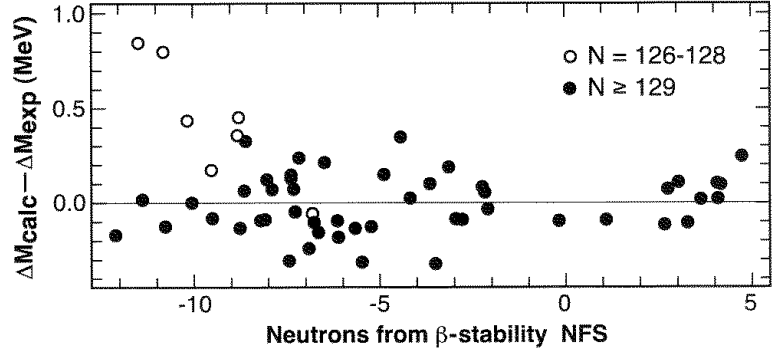} \end{center}
\caption{Deviations of the predicted masses,
    eq.~(\ref{eq1}), from the data for the 56 new masses
    measured after the adjustments were made. Shown as function of
    Neutrons From Stability (NFS).}
\end{figure}

\begin{figure}
\begin{center}
\leavevmode \epsfysize=12.0cm \epsfbox{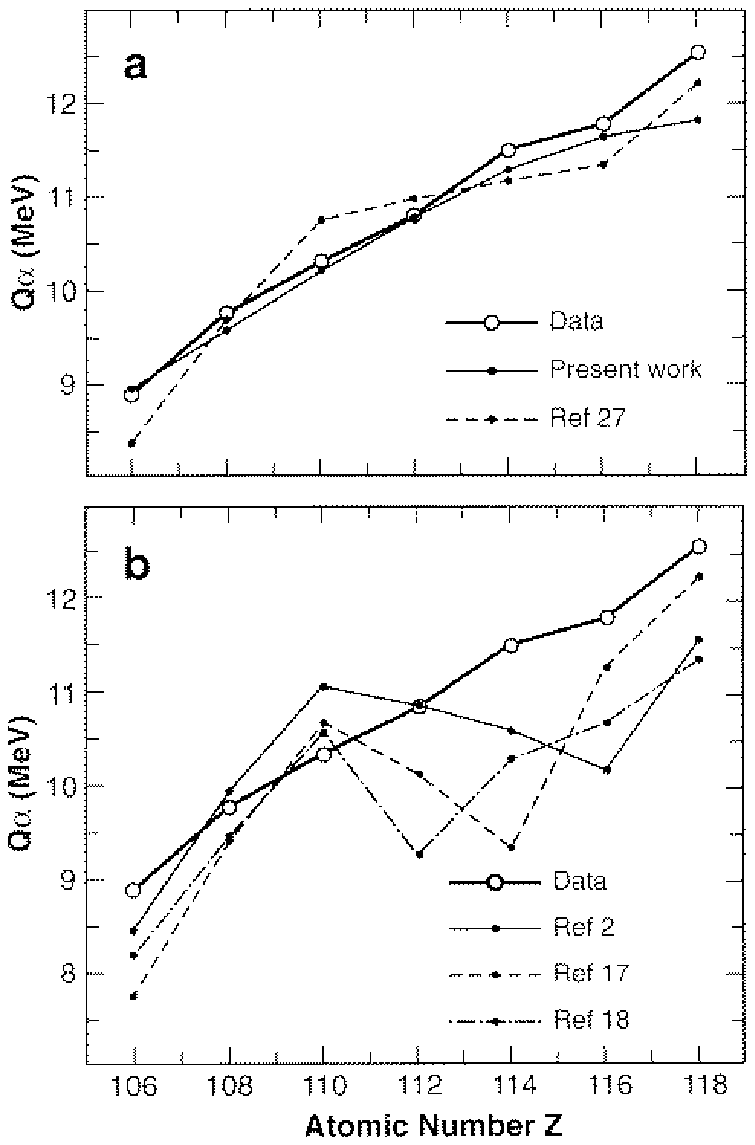} \end{center}
\caption{Experimental [1] and predicted Q$_{\alpha}$ values of
  the $^{293}$118 decay chain. \\ (a) Predictions of
 eq.~[1] and of ref. [27]. \\
  (b) Predictions of refs. [2,17,18].}
\end{figure}


\begin{thebibliography}{99}
    \bibitem{nin99}  V. Ninov et al., Phys. Rev. Lett. {\bf 83}, 1104
    (1999).

    \bibitem{cnh99}  S. \'{C}wiok, W. Nazarewicz and P.H. Heenen, Phys.
    Rev. Lett. {\bf 83}, 1108 (1999).

    \bibitem{mon94}  P. M\"oller and J.R. Nix, J. Phys. {\bf G20}, 1681
    (1994).

    \bibitem{smo97}  R. Smola\'{n}czuk, Phys. Rev. {\bf C56}, 812 (1997).

    \bibitem{zam95}  N.V. Zamfir et al., Phys. Lett. {\bf B357}, 515
    (1995).

    \bibitem{zel98}  N. Zeldes, Phys. Lett. {\bf B429}, 20 (1998).

    \bibitem{mys98}  W.D. Myers and W.J. Swiatecki, Phys. Rev. {\bf
    C58}, 3368 (1998).

    \bibitem{cwi96}  S. \'{C}wiok et al., Nucl. Phys. {\bf A611}, 211
    (1996).

    \bibitem{rut97}  K. Rutz et al., Phys. Rev. {\bf C56}, 238 (1997).

    \bibitem{ben99}  M. Bender et al., Phys. Rev. {\bf C60}, 034304
    (1999).

    \bibitem{kru00}  A.T. Kruppa et al., Phys. Rev. {\bf C61}, 034313
    (2000).

    \bibitem{liz76}  S. Liran and N. Zeldes, At. Data Nucl. Data
    Tables {\bf 17}, 431 (1976).

    \bibitem{zel96}  N. Zeldes, in Handbook of Nuclear Properties,
    edited by D.N. Poenaru and W. Greiner (Clarendon Press, Oxford,
    1996), p. 12.

    \bibitem{liran}  S. Liran, Calculation of Nuclear Masses in the
    Shell Model, Ph.D. Thesis, Jerusalem, 1973. (In Hebrew.
    Unpublished.)

    \bibitem{hau84}  P.E. Haustein in Atomic Masses and Fundamental
    Constants 7, edited by O. Klepper, THD-Schriftenreihe Wissenschaft
    und Technik, Bd. 26 (Darmstadt, 1984), p. 413.

    \bibitem{mol95}  P. M\"oller et al., At. Data Nucl. Data Tables {\bf
    59}, 185 (1995).

    \bibitem{mnk97}  P. M\"oller, J.R. Nix and K.-L. Kratz, At. Data Nucl.
    Data Tables {\bf 66}, 131 (1997).

    \bibitem{abo95}  Y. Aboussir et al., At. Data Nucl. Data Tables {\bf
    61}, 127 (1995).

    \bibitem{oga99}  Yu.Ts. Oganessian et al., Phys. Rev. Lett. {\bf 83},
    3154 (1999).

    \bibitem{wag71}  A.H. Wapstra and N.B. Gove, Nuclear Data Tables
    {\bf 9}, 265 (1971).

    \bibitem{auw95}  G. Audi and A.H. Wapstra, Nucl. Phys. {\bf A595},
    409 (1995).

    \bibitem{zel67}  N. Zeldes, Ark. Fys. {\bf 36}, 361 (1967).

    \bibitem{com70}  E. Comay et al. in Int. Conf. on the Properties of
    Nuclei Far from the Region of Beta-Stability, CERN 70-30 (Geneva,
    1970) p. 165.

    \bibitem{mys96}  W.D. Myers and W.J. Swiatecki, Nucl. Phys. {\bf
    A601}, 141 (1996).

    \bibitem{duz95}  J. Duflo and A.P. Zuker, Phys. Rev. {\bf C52}, R23
    (1995).

    \bibitem{zgl65}  N. Zeldes, M. Gronau and A. Lev, Nucl. Phys. {\bf
    63}, 1 (1965).

    \bibitem{smo99}  R. Smola\'nczuk, Phys. Rev. {\bf C60}, 021301 (1999).

    \bibitem{ben00}  M. Bender, Phys. Rev. {\bf C61}, 031302 (R) (2000).

\end{thebibliography}
\end{document}